\documentclass[aps,pre,twocolumn,nofootinbib,floatfix]{revtex4}
\usepackage{amsmath}
\usepackage{epsfig}

\input{tcilatex}

\begin{document}

\title{Orientations of two coupled molecules}
\author{Y. Y. Liao, Y. N. Chen, and D. S. Chuu\thanks{%
Corresponding author email address: dschuu@cc.nctu.edu.tw;
Fax:886-3-5725230; Tel:886-3-5712121-56105.}}
\affiliation{Department of Electrophysics, National Chiao-Tung University, Hsinchu 300,
Taiwan}
\date{\today }

\begin{abstract}
Orientation states of two coupled polar molecules controlled by laser pulses
are studied theoretically. By varying the period of a series of periodically
applied laser pulse, transition from regular to chaotic behavior may occur.
Schmidt decomposition is used to measure the degree of entanglement. It is
found that the entanglement can be enhanced by increasing the strength of
laser pulse.

\end{abstract}

\maketitle

\address{Department of Electrophysics, National Chiao Tung University,
Hsinchu 300, Taiwan}





The ability to control the alignment of molecules may have important purpose
in stereodynamics \cite{1}, surface catalysis \cite{2}, trapping molecules %
\cite{3}, molecular focusing \cite{4}, and nanoscale design \cite{5,6}.
Experimentally, several methods have been used to control the orientations
of molecules \cite{7,8,9,10,11}. For example, by turning on a picosecond
laser pulse adiabatically, the pendular states-- hybrid of field-free
molecular eigenstates \cite{12,13,14} --can be created. A femtosecond laser
pulse, like impulsive excitation, is found to be able to generate a
field-free orientation \cite{15,16}.

Since entangled states are fundamental for quantum information processing %
\cite{17,18}, many research works have been proposed to generate
entanglement in quantum-optic and atomic systems \cite{19,20}. In this
Letter we propose a novel way to generate entanglement between two coupled
identical polar molecules separated in a distance of tens of nanometers.
Both molecules are irradiated by ultra-short pulses of laser light. Our
study shows the entanglement induced by the dipole interaction can be
enhanced by controlling the inter-molecule distance and the field strength
of laser pulse.

Consider now two identical polar molecules (separated by a distance of $R$
). There exists dipole-dipole interaction between these two molecules.
Ultrashort half-cycle laser pulses are applied to both molecules \cite{21}.
The Hamiltonian of the system can be written as%
\begin{equation}
H=\sum_{j=1,2}\frac{\hbar ^{2}}{2I}L_{j}^{2}+U_{dip}+H_{l},
\end{equation}%
where $L_{j}^{2}$ and\ $\frac{\hbar ^{2}}{2I}$ $\left( =B\right) $ are the
angular momentum operator and rotational constant, respectively. $U_{dip}$ $%
=[\vec{\mu}_{1}\cdot \vec{\mu}_{2}-3\left( \vec{\mu}_{1}\cdot \widehat{e}%
_{R}\right) \left( \vec{\mu}_{2}\cdot \widehat{e}_{R}\right) ]/R^{3}$ is the
dipole interaction between two molecules, where $\vec{\mu}_{1}$ and $\vec{\mu%
}_{2}$ are the dipole moments.\ For simplicity, the dipole moments of two
molecules are assumed to be identical, i.e. $\mu _{1}=\mu _{2}=\mu $. The
field-molecule coupling can be expressed as $H_{l}=-\mu E\left( t\right)
\cos \theta \cos \left( \omega t\right) -\mu E\left( t\right) \cos \theta
^{^{\prime }}\cos \left( \omega t\right) ,$where $\theta $\ and $\theta
^{\prime }$\ are the angles between the dipole moments and laser field. The
laser profile is assumed to be Gaussian shape, i.e. $E\left( t\right)
=E_{0}e^{-\frac{(t-to)^{2}}{\sigma ^{2}}},$ where $E_{0}$ is the field
strength, $t_{0}$ is the center of peak, and $\sigma $ is the pulse
duration. With these assumptions, the time-dependent Schr\"{o}dinger
equation can be solved by expanding the wave function in terms of a series
of field-free spherical harmonic\ functions 
\begin{equation}
\Psi =\sum_{lml^{\prime }m^{\prime }}c_{lml^{\prime }m^{\prime }}\left(
t\right) Y_{lm}\left( \theta ,\phi \right) Y_{l^{\prime }m^{\prime }}\left(
\theta ^{\prime },\phi ^{\prime }\right) ,
\end{equation}%
where $\left( \theta ,\phi \right) $ and $\left( \theta ^{\prime },\phi
^{\prime }\right) $ are the coordinates of first and second molecule
respectively. $c_{lml^{\prime }m^{\prime }}\left( t\right) $ are the
time-dependent coefficients corresponding to the quantum numbers $\left(
l,m;l^{\prime },m^{\prime }\right) $ and can be determined by solving the
Schr\"{o}dinger equations numerically. In above equation, total wavefunction
has no spatial dependence since we keep inter-molecule separation $R$ as a
fixed parameter. One might argue that variation of $R$ is inevitable because
of the effects of laser fields or inter-molecule vibrations. However, recent
experiments have shown that it is possible to resolve two individual
molecules separated by tens of nanometers when they are hindered on a
surface \cite{22,23}. In principle, the free orientation model can be easily
generalized to hindered ones by replacing the spherical harmonic\ functions
with hindered wavefunctions as shown in our previous work \cite{24}. The
essential physics discussed below should be similar.

The orientations $\left\langle \cos \theta \right\rangle $ and $\left\langle
\cos \theta ^{\prime }\right\rangle $ can be evaluated immediately after the
coefficients $c_{lml^{\prime }m^{\prime }}\left( t\right) $ are determined.
The parameters for numerical calculations are based on NaI molecule whose
dipole moment is 9.2 debyes and rotational constant is 0.12 cm$^{-1}$ in the
ground state. Other parameters are referred to Ref. \cite{16}. The duration
and frequency of the half-cycle pulse are set to 279 fs and 30 cm$^{-1}$,
respectively. The center of peak is 1200 fs and the initial condition is $%
c_{0000}\left( t=0\right) =1$. Fig. 1 shows the orientations of the first
and second molecules after a \emph{single} laser pulse is applied on both
molecules. For $R=3\times 10^{-8}$ m, the behavior of the first molecule is
quite close to that of a free rotor \cite{16}. This is not surprising
because the dipole interaction is weak for this molecule separation.
However, as two molecules get close enough (Fig. 1 (b)), both molecules
orient disorderly, and the periodic behavior disappears. This is because the
dipole interaction is increased as the distance between the molecules is
decreased, and the energy exchange between two molecules becomes more
frequently. The regular orientation caused by the laser pulse is inhibited
by the mutual interaction.

The populations of some low-energy levels are shown in the lower panels of
Fig. 1(a) and (b). The solid, dashed, and dotted lines represent the
populations of the states (1,0;0,0), (1,0;1,0), and (2,0;1,0), respectively.
These states show different degrees of periodic behavior at different
distances. However, the populations of some higher excited states, for
example the (3,0;1,0) state in the inset of Fig. 1(b), display different
degrees of irregularity. This manifests a fact that the nonlinear effect,
caused by the reduction of $R$, does not affect the regularity of the
low-lying states, and the origin of the irregularity is caused by the higher
excited states. 
\begin{figure}[th]
\includegraphics[width=7.5cm]{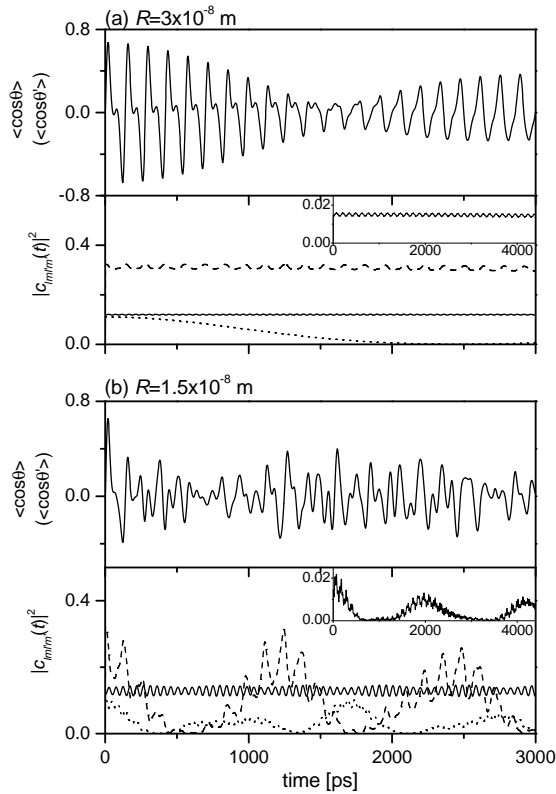}
\caption{Upper panels of Fig. 1(a) and (b) show the orientations of the two
molecules at different distances. Lower panels: The populations of the
states $\left( l,m;l^{\prime },m^{\prime }\right) $=$\left( 1,0;0,0\right) $
(solid lines), $\left( 2,0;1,0\right) $ (dotted lines), $\left(
1,0;1,0\right) $ (dashed lines). The insets in (a) and (b) represent the
population of state $\left( 3,0;1,0\right) .$ }
\end{figure}

Consider now the molecules are irradiated by a series of laser pulses
periodically. As shown in Fig. 2 (a), if the period of the applied
periodically laser pulse $T$ \ is equal to $\hbar /B$ , then both molecules
behave disorderly no matter how the distance $R$ is varied. The chaotic
behavior of the molecules can be ascribed to the well-known
''kicked-rotor''problem. However, a series of \emph{regular-like}
orientations marked by dotted and dashed lines are present in Fig. 2 (b) if $%
T$ is equal to $\pi \hbar /B$. \ For a free rotor under a \emph{single}
kick, this interesting phenomenon comes from the situation as the magnitude
of the orientation returns to its initial condition ($\left\langle \cos
\theta \right\rangle =0$) after a certain period $T$ \cite{16}. Therefore,
for two molecules in weak interaction limit ($R=3$ $\times $ $10^{-8}$ m),
the wavepacket-like orientation is similar to that of a single free rotor
under the same laser period. The difference is the suppression of the
amplitudes at long time (dashed lines). It means that the dipole force can
generate some accidental phases to perturb the regularity of the coupled
system. The lower panel of Fig 2 (b) exhibits that the suppression of the
regularity is quicker if the dipole force is stronger. 
\begin{figure}[th]
\includegraphics[width=7.5cm]{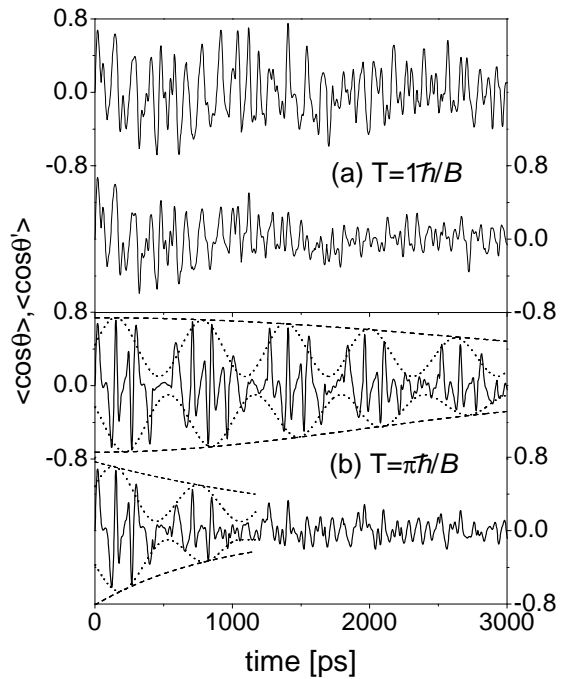}
\caption{The orientations of the first and second molecules under periodic
laser pulses with the intensity $E_{0}=$ $3$ $\times $ $10^{7}$ V/m and
periods T$=$ (a) $1\hbar /B,$ (b) $\protect\pi \hbar /B$ ps. The upper and
lower panels of (a) and (b) correspond to the distances $R=3$ $\times $ $%
10^{-8}$ and $2$ $\times $ $10^{-8}$ m, respectively. }
\end{figure}

Let us now turn our attention to the entanglement generated in this system.
The coupled molecules in fact can be expressed as a pure bipartite system 
\begin{equation}
\left| \Psi \right\rangle =\sum_{lml^{\prime }m^{\prime }}c_{lml^{\prime
}m^{\prime }}\left( t\right) \left| Y_{lm}\right\rangle \left| Y_{l^{\prime
}m^{\prime }}\right\rangle ,
\end{equation}%
The \emph{partial} density operator for the first molecule is given by 
\begin{equation}
\rho _{\text{mol 1}}=\text{Tr}_{\text{mol 2}}\left| \Psi \right\rangle
\left\langle \Psi \right| .
\end{equation}%
Following the procedure of Schmidt decomposition, the bases of molecule 1 is
rotated to make the reduced density matrix $\rho _{\text{mol 1}}$ diagonal.
The entangled state can then be represented by a biorthogonal expression
with positive real coefficients as given by%
\begin{equation}
\left| \Psi \right\rangle =\sum_{lm}\sqrt{\lambda _{lm}}\left|
Y_{lm}\right\rangle _{\text{mol 1}}\left| Y_{lm}\right\rangle _{\text{mol 2}%
},
\end{equation}%
where $\lambda _{lm}$ is the eigenvalue corresponding to $\left|
Y_{lm}\right\rangle _{\text{mol 1}}\left| Y_{lm}\right\rangle _{\text{mol 2}}
$. The measure of entanglement for the coupled molecules can be parametrized
by the von Neumann entropy 
\begin{equation}
\text{Entropy}=-\sum\limits_{lm}\lambda _{lm}\text{log}_{\text{n}}\lambda
_{lm}.
\end{equation}%
Fig. 3 shows the time-dependent entropy after one pulse passes through this
system. For inter-distance $R=5$ $\times $ $10^{-8}$ m, the entropy
increases slowly from zero. For $R=1.5$ $\times $ $10^{-8}$ m, on the
contrary, the entropy grows rapidly with the increasing of time because the
dipole force is stronger. Notes that the entropy only varies within a finite
range at long time regime. This indicates that the system reaches a dynamic
equilibrium state even though the dipole force is still present. 
\begin{figure}[th]
\includegraphics[width=7.5cm]{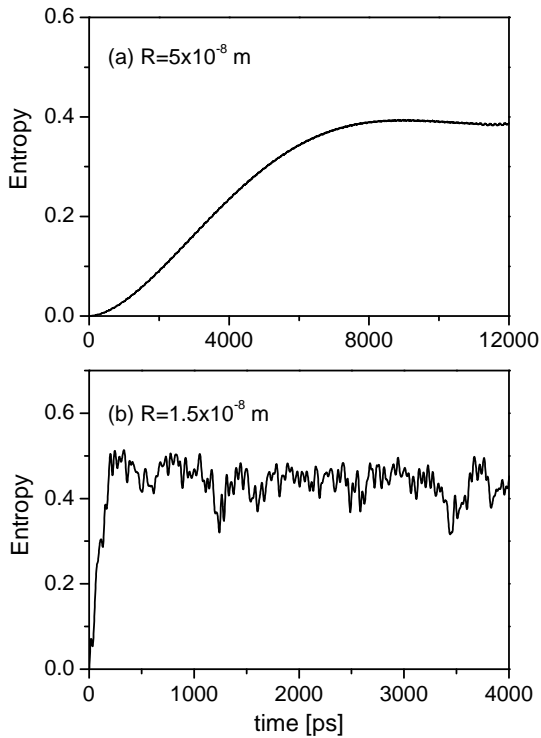}
\caption{Time evolution of the entropy after applying single laser pulse for
(a) $R=5$ $\times $ $10^{-8}$ m and (b) $R=1.5$ $\times $ $10^{-8}$ m. The
field strength is set equal to $3\times 10^{7}$ V/m.}
\end{figure}

Fig. 4 (a) illustrates the variations of the entropy with respect to
different field strengths of the applied laser pulse as $R$ is set equal to $%
1.5$ $\times $ $10^{-8}$ m. For the field strength $E_{0}=1.5\times 10^{7}$
V/m, an irregular-like behavior of the entropy is obtained, and its value is
not large enough for quantum information processing. However, Fig. 4 (b)
shows that the degree of entanglement can be enhanced if one increases the
field strength. This can be understood well by studying the relationship
between the dipole interaction and the field strength. If the effect of
dipole interaction overwhelms the laser field, most of the populations are
distributed on the low-lying states. In this case, the entropy from Schmidt
decomposition is certainly small as shown in Fig. 4 (a). On the other hand,
if the field strength plays a dominant role, the distribution of molecular
states covers a wider range and the entropy is enhanced in this limit. 
\begin{figure}[th]
\includegraphics[width=7.5cm]{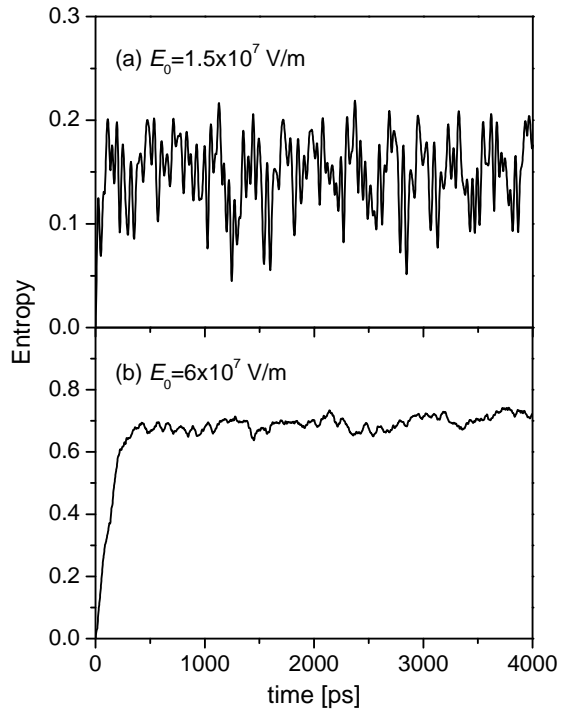}
\caption{Time evolution of the entropy for inter-molecule separation $R=1.5$ 
$\times $ $10^{-8}$ m. The degree of entanglement can be enhanced if one
increases the field strength.}
\end{figure}
\ \ \ 

A few remarks about the differences between present proposal and previous
works on generating entangled states should be emphasized here. In our
model, the rotational excited states instead of internal \emph{electronic}
states of the molecules are considered. Second, the laser frequency in our
work is tuned far-away from resonance, while conventional creation of
entanglement depends on the resonant driving pulses. This means our work
provides a wider range to select the laser frequency to create entanglement.
As for the effect of decoherence, our entanglement is formed by the excited
rotational states, instead of the vibrational states. Therefore, the
decoherence is dominated by photon emission even if the molecules are
attached to the surface of a solid.

In conclusion, we have studied a system of two coupled polar molecules,
irradiated by laser pulse. The chaotic behavior of the orientations comes
from the populations in the excited states of higher mode numbers. The
entangled states vary with the field strength of laser pulse. And more
stable entangled state can be generated even though the coupled system has
irregular orientations. This feature may be useful in quantum information
processing.

We would like to thank to C. M. Li and Dr. C. H. Chang at NCTS for helpful
discussions. A critical reading of the paper by Dr. J. Y. Hsu at NCHC is
greatly acknowledged. This work is supported by the National Science
Council, Taiwan under the grant numbers NSC 92-2120-M-009-010 and NSC
93-2112-M-009-037.


\begin{thebibliography}{99}
\bibitem{1} Special issue on Stereodynamics of Chemical Reaction [J. Phys.
Chem. A \textbf{101}, 7461 (1997)].

\bibitem{2} V. A. Cho and R. B. Bernstein, J. Phys. Chem. \textbf{95}, 8129
(1991).

\bibitem{3} B. Friedrich, Phys. Rev. A \textbf{61}, 025403 (2000).

\bibitem{4} H. Stapelfeldt, H. Sakai, E. Constant, and P. B. Corkum, Phys.
Rev. Lett. \textbf{79}, 2787 (1997).

\bibitem{5} T. Seideman, Phys. Rev. A \textbf{56}, R17 (1997).

\bibitem{6} R. J. Gordon, L. Zhu, W. A. Schroeder, and T. Seideman, J. Appl.
Phys. 94, 669 (2003).

\bibitem{7} W. Kim and P. M. Felker, J. Chem. Phys. \textbf{104}, 1147
(1996).

\bibitem{8} W. Kim and P. M. Felker, J. Chem. Phys. \textbf{108}, 6763
(1998).

\bibitem{9} H. Sakai, C. P. Safvan, J. J. Larsen, K. M. Hilligs\o e, K.
Hald, and H. Stapelfeldt, J. Chem. Phys. \textbf{110}, 10235 (1999).

\bibitem{10} J. J. Larsen, H. Sakai, C. P. Safvan, I. Wendt-Larsen, and H.
Stapelfeldt, J. Chem. Phys. 111, 7774 (1999).

\bibitem{11} H. Sakai, S. Minemoto, H. Nanjo, H. Tanji, and T. Suzuki, Phys.
Rev. Lett. \textbf{90}, 083001 (2003).

\bibitem{12} B. Friedrich and D. R. Herschbach, Phys. Rev. Lett. \textbf{74}%
, 4623 (1995).

\bibitem{13} L. Cai, J. Marango, and B. Friedrich, Phys. Rev. Lett. \textbf{%
86}, 775 (2001).

\bibitem{14} B. Friedrich and D. Herschbach, J. Chem. Phys. \textbf{111},
6157 (1999).

\bibitem{15} I. Sh. Averbukh and R. Arvieu, Phys. Rev. Lett. \textbf{87},
163601 (2001).

\bibitem{16} M. Machholm and N. E. Henriksen, Phys. Rev. Lett. \textbf{87},
193001 (2001).

\bibitem{17} C. H. Bennett and D. P. DiVincenzo, Nature (London) \textbf{404}%
, 247 (2000).

\bibitem{18} Y. N. Chen, D. S. Chuu, and T. Brandes, Phys. Rev. Lett. 
\textbf{90}, 166802 (2003).

\bibitem{19} J. I. Cirac and P. Zoller, Phys. Rev. Lett. \textbf{74}, 4091
(1995).

\bibitem{20} D. DeMille, Phys. Rev. Lett.\textbf{\ 88}, 067901 (2002).

\bibitem{21} D. You, R. R. Jones, P. H. Bucksbaum, and D. R. Dykaar, Opt.
Lett. \textbf{18}, 290 (1993).

\bibitem{22} C. Hettich, C. Schmitt, J. Zitzmann, S. Kuhn, I. Gerhardt, and
V. Sandoghdar, Science \textbf{298}, 385 (2002).

\bibitem{23} S. Yasutomi, T. Morita, Y. Imanishi, and S. Kimura, Science 
\textbf{304}, 1944 (2004).

\bibitem{24} Y. T. Shih, Y. Y. Liao, and D. S. Chuu, Phys. Rev. B \textbf{68}%
, 075402 (2003).
\end{thebibliography}
\end{document}